# ON EXPLORING CONSUMERS' TECHNOLOGY FORESIGHT CAPABILITIES
## *An analysis of 4,000 mobile service ideas*

Petteri Alahuhta, Pekka Abrahamsson
*VTT Technical Research Centre of Finland, Kaitoväylä 1, Oulu, Finland*
*petteri.alahuhta@vtt.fi, pekka.abrahamsson@vtt.fi*

Antti Nummiaho
*VTT Technical Research Centre of Finland, Vuorimiehentie 3, Espoo, Finland*
*antti.nummiaho@vtt.fi*

Abstract: Lead user driven innovation and open innovation paradigms seek to involve consumers and common people to innovative product development projects. In order to help developers choose ideas that meet the end users' needs, we undertook a massive collaborative research effort and collected 40,000 ideas from 2,150 common people about future mobile services that they would like to use. We inspired each people to produce tens of mobile service ideas. In this paper we carry out an analysis for 4,000 ideas from the idea database. We had a particular interest in whether peoples' ideas can be used in foreseeing the technology development needs. The results show that end users produce ideas that are conservative more than novel. Therefore, we claim that consumers' technology foresight horizon is limited by the existing technological base. The second finding, linked to the previous one, is that the great majority of the ideas that consumers expressed could be realised utilizing existing technologies. The implication of this finding is that the idea database should be an interesting source of ideas for service developers. The third finding of the study, related to the methodology, is that a vast number of ideas can be collected fairly easily but analyzing them cost effectively is a challenge.

## 1 INTRODUCTION

End users of the products and services are instrumental for the success of any innovation. They will determine which products and services will ultimately become successful and which ones will fail. The end users' market behaviour is difficult to forecast, and indeed history has shown that markets and end users may take unexpected turns. As an example, in the mobile service segment the high popularity of SMS (Short Message Service) led the operators to forecast that an MMS (Multimedia Messaging Service) offering would become an instant hit among mobile phone users as soon as the new technology was made available. For several reasons, the adoption rate of MMS has been significantly lower than the SMS service and in majority of the mobile service markets the SMS still keeps dominating the service offering despite of the opportunities that the newer technology offers.

Von Hippel (1986, 2005) has been one of the first authors to promote the use of consumers' technological foresight capabilities to the fullest extent. In his seminal work on lead user driven innovation in 1986, he argues that in the area of high-tech product development lead users are actually ahead of the market place and possibly possess innovative ideas about the product or technology that they use. Von Hippel (2005) argues also that lead users may be quite willing to share their product related innovations freely and that these ideas are often commercially attractive.

Chesbrough (2003) popularized the concept of open innovation to distinguish from the traditional model of innovation, which he calls a closed innovation model. Chesbrough (2003) argues that external R&D (Research and Development) can create significant value and the internal R&D is needed to claim a part of this value. His open innovation model relies on building the business model first. He argues that companies should learn to increase the effective usage of a company's IP (Intellectual Property) either by selling it or offering it outside of the company's boundaries by other means. Also, learning to acquire IP from companies, individuals and user communities (Baldwin, Hienerth & Von Hippel 2006) outside, whenever it fits the business model, bears significance in producing innovative products and services. In Chesbrough's (2003) terms the open innovation model forms a new imperative for creating and profiting from technology.

Based on Von Hippel's and Chesbrough's concepts of lead user driven innovation and open innovation, the involvement of consumers in the development of future mobile services would seem to be of paramount importance to guarantee the continuous feed of innovative products and services. Little is known, however, to what extent the end users are capable of forecasting into the future (Un, Price 2007). Thus, the potential technological foresight capability of a mobile terminal user remains an area of research with little results.

When asked about future technology, we have a tendency to rely on forecasts made by information and technology research companies such as Gartner (www.gartner.com) and Standish Group (www.standishgroup.com/). Their predictions on future service development, builds upon technology. Markets and end-users are known to take unexpected turns. They may be even reluctant to adopt new technologies or services. behaviourTherefore it can be asked, how accurately are technology research companies able to predict future services, as the diffusion of services is dominated largely by other factors than those related to any particular technology.

In order to help companies and developers meet the needs of the growing number of mobile terminal users, we undertook a massive collaborative research effort and collected 40,000 ideas from 2,150 ordinary people about future mobile services that they would like to use. We inspired people not to think about the technology, but to express their concrete needs. Moreover, we did not request a single or few ideas from these people. Rather, we motivated people to come up with 30 or more ideas even if they would feel them to be either "silly" or not realistic in their minds.

In this paper, we report the results from a study where 4,000 ideas from the database were chosen for a technical scrutiny. We are particularly interested in understanding the technical foresight capability of an end user. We also wanted to know if by analyzing users' ideas, one could identify specific technology development needs. Therefore, we paid particular attention to identifying possible technology related obstacles in front of successful mobile services, such as limitations in human technology interaction or in communication capabilities. As context awareness has been advertised as a technology to bridge the shortcomings of the mobile HCI (Cheverst et al. 2000, Korpipaa et al. 2006), we also analyzed to what extent the ideas suggest that this would be a desired development trend.

The paper is organized as follows: the next section outlines the research design of the study focusing on research methods, data collection and analysis. Also, the outline of the Idea Movement initiative that the present researchers took alongside with the principal high level results of the Idea Movement itself is presented. This is followed by the results of the analysis and a brief discussion of the implications for the mobile service business and research. The paper is concluded with final remarks.

## 2 RESEARCH DESIGN

This section discusses the issues related to research design. We will first introduce the research method and settings, and then we will briefly describe the data collection and means of analysis of the material. Finally, we will give some examples of collected ideas stored in the database.

### 2.1 Research method and settings

This study is an explorative research on the mobile service needs. The phases of an innovation process have been discussed in the literature (e.g., Koen et al. 2002). There is a strong need to better understand the role and potential of end users in foreseeing the technology development needs. The problem has also been discussed by Un & Price (2007). In order to explore the possibilities of ideas, we launched a national research project called the Idea Movement. The project started in the beginning of 2006. We set the goal to collect at least 35,000 ideas for mobile applications straight from the citizens, and to make these ideas accessible for everyone by publishing them in the Internet. Giving companies, organizations and individuals the opportunity to build on the ideas of thousands of people, we aim to accelerate the development and commercialization of new mobile services.

### 2.2 Data collection and analysis

Ideas were collected systematically in 31 workshops and events organized around Finland. Altogether 2,150 people participated in the events. The majority of the participants were university students, schoolchildren and elderly people. Workshops were also organized in schools, at workplaces and even in a shopping centre. Furthermore, there was a possibility to post ideas on the Idea Movement website, and about 200 people communicated their ideas in this way. Table 1 summarizes the division of ideas according to different age groups.

Table 1. Distribution of ideas in the Idea database.

| Age group (years) | number of ideas | Proportion of ideas |
|---|---|---|
| 7-12 | 1,800 | 4% |
| 13-16 | 6,500 | 16% |
| 17-19 | 4,300 | 10% |
| 20-50 | 21,000 | 51% |
| 50-90 | 4,300 | 10% |
| Unclassified | 3,100 | 8% |
| Total | 41,000 | 100% |

Each workshop was kicked off with a short introduction to idea generation techniques, followed by brainstorming sessions both individually and in groups. At the beginning of each event we asked the participants to produce ideas about a given topic, but we also accepted ideas that did not match the original subject of the particular brainstorming session. We instructed each participant to produce 20 ideas individually and then form groups of 3-4 people. The ultimate goal of idea generation was that each group would deliver together 100 ideas, or more.

Ideas are 1-2 sentence descriptions of a mobile service idea or an expression of a need that they think could be fulfilled utilizing mobile technology. The progress of a brainstorming session has been described in more detail by Leikas (2007).

Idea analysis was carried out in two phases. First a group of five reviewers did a qualitative analysis of 2,000 ideas so that we got an understanding of the potential of the ideas. According to the reviewers' subjective opinion, each idea was classified into one of the following four categories: *Excellent*, *Interesting*, *Conventional* and *Not a mobile service idea.*

We also carried out an analysis of the technical aspects of the collected ideas in order to get an understanding of the technologies that are required for implementing these ideas. Also, the essential constraints that may currently prevent the implementation were considered.

For the analysis three technology areas (access technology, context information, human-technology interaction) and major technical constraints were identified. Furthermore, we explored what kinds of integrated or interconnected devices were proposed to be part of mobile services. The analysis was carried out by reviewing a selected portion of the ideas and categorizing them based on the defined categories. Each idea was classified based on its most obvious way of implementation. For this paper, we selected 4,000 ideas (i.e., 10% of all ideas) for a detailed inspection. The selected dataset consists of 2,000 ideas produced by high-school students (the age group of 17-18 years) and 2,000 ideas from university students and employees (the age group of 20-50 years).

### 2.3 Data examples from the idea base

The themes of covered a large variety of different topics including *Public transportation, Travelling, Work and Learning, Culture and entertainment, Hobbies, Wellbeing and health, Shopping and service, Family, Friends and relatives, Household management and living* and *Everyday activities.*

In order for the reader to comprehend the nature and type of the mobile service ideas, we have listed below some ideas classified to the four suggested categories: *Excellent ideas, Interesting ideas, Conventional ideas* and *Not a mobile service idea.* These categories were created in an early phase of the analysis in order to quickly have an exploratory view of the ideas in the database.

Examples of ideas classified as *Excellent* are*:*
- Warning if parking time is running out. Possibility to automatically buy or order additional parking time.
- Location-based filtering of incoming calls. E.g., No work-related calls at home.
- Locating of friends (if they allow).

Examples of *Interesting ideas* are*:*
- In a rally, one can get information about the condition of a car. A spectator will know if the car is going to break down.
- Service that tells you how strong the punch in your glass is.
- Opportunity to order "good explanations" when coming home late.

Some ideas classified as *Conventional are:*
- The controlling of a home automation system by a mobile handset. Being able to switch on/off lights, heating, etc.
- Message from library if a new book by your favourite author is available.
- Service for getting current hit music to your mobile phone.

Finally, examples of ideas that are *not a mobile service idea* include*:*
- Mobile phone making a cup of coffee
- 3" nails inside the mobile phone
- Rat trap

## 3 RESULTS

In this section we will describe the results of the technical analysis of 4,000 ideas. At first the application categories are introduced, followed by technical constraints in five major areas suggested by the literature.

### 3.1 Application categories

We first classified mobile service ideas into eight distinct categories based on the idea clusters identified in the data and had one category for miscellaneous ideas. The results have been summarized in Table 2.

Table 2. Distribution of ideas in application categories

| Category | Description | Total |
|---|---|---|
| Information pull | Retrieving information for some purpose (possibly based on location). | 30% |
| Information push | Receiving information automatically (possibly based on location). | 14% |
| Locating (persons / objects) | Locating or following some (nearest) person or object. | 9.2% |
| Communication | Social discussion channel. | 7.0% |
| Service request | Ordering a personal service (possibly based on location). | 4.7% |
| Content production | Producing content. | 4.3% |
| Payment | Using mobile device as a means of payment. | 4.0% |
| Identification | Using mobile device as a identification device. | 3.4% |
| Others | Applications that do not fit into other categories. | 24% |

According to technical analysis people see mobile services largely as an information channel, which can be used when ever they need to know something. 30% of ideas fall into this category. muchSignificantly fewer ideas proposed automatic information delivery to users' terminals. Only 14% of the ideas proposed push services such as advertisement. Locating missing objects and persons representrepresented 9.2% and communicating, for example, with social communities 7% of ideas. Service requests (4.7%), mobile content production (4.3%), payment (4%) and identification (3.4%) representrepresented a perhaps surprisingly low popularity in the analyzed set of ideas.

24% of the ideas fell into the category *others*. We think this indicates that there is a versatile set of activities in everyday life where mobile technology could provide value to the users.

### 3.2 Technical constraints

In this paper we want to explore the nature of user innovation ideas from the technical viewpoint. We are particularly interested in what kinds of technologies are needed to realise the ideas that users have proposed. More specifically we are interested in

a) what seems to be the main constraints for services not being developed,
b) what communication and access technologies are required for the ideas,
c) what kind of context information, if any, is required for services that users propose,
d) what are the challenges in Human Technology Interaction (HTI) technologies, and
e) what kinds of integrated or interconnected devices are required for the ideas.

#### 3.2.1 Analysis of major constraints for adoption of mobile services

One of the main findings of our technical analysis is that there are no major technical constraints in developing most of the mobile service ideas into actual mobile services for people. However, we can pinpoint some constraints that are quite typical for mobile services and well known by the literature.

Even if these constraints are not immediate road blocks for services, they weaken the user experience so much that eventually users may not start using the services in the first place or they stop using the services because of the poor experience.

In Table 3 technical factors that may hinder user experience have been summarized. Small screen size and low bandwidth are perhaps the most important constraints in mobile services. Other factors that make it difficult to realise some of the ideas are low processing power, high power consumption of the terminals and a limited amount of memory available in the terminal.

Some of these constraints will be overcome in time. Such , such as bandwidth, processing power and amount of memory. Screen size and power consumption, however, are harder obstacles for developers. The trend towards extensive multimedia communication requires as large a screen as possible and higher processing power that leads to higher power consumption.

Table 3. Major technical constraints identified in collected mobile service ideas

| Category | Description | Total |
|---|---|---|
| Screen size | The user interface requires a large screen. | 3.5% |
| Bandwidth | A lot of network traffic occurs. | 2.9% |
| Processing power | Processing power is especially important. | 1.0% |
| Amount of memory | A lot of multimedia or other memory expensive data is handled. | 0.96% |
| Battery duration | A lot of power is needed. | 0.91% |
| Touch screen | The user should be enabled to interact with the application by touching pictures or words on the screen. | 0.32% |
| Keyboard | The user must type lots of text. | 0.30% |

It is also to note that the total percentage of ideas having major technical constraints is very low. In other words, technology is not the main constraint in implementing most of the ideas.

### 3.2.2 Analysis of required communication and access technologies

In the technical analysis we wanted to consider what communication and/or service access technologies are required for realizing the ideas. Table 4 summarizes the analysis of required access technologies.

In the table we can see that the vast majority (67%) of foreseen services can be realizedrealised utilizing existing access technologies such as GSM/GPRS (2G) or 3$^{rd}$ generation cellular networks (3G). 13% of ideas required short-range communication solutions, such as Bluetooth or Near Field Communication (NFC). Quite few services really required such communication technologies as satellite communication, cell-casting or a combination of different communication technologies.

During the time of collecting the ideas (in 2006) there were quite lively debates about the promised break-through of mobile-TV. Our study did not support the need of these technologies and it seems that users have not adopted to broadcasting-type access technologies beyond FM-radio-receivers in their mobile handsets.

Table 4. Required access technologies

| Category | Description | Total |
|---|---|---|
| 3G | Text and multimedia based services. | 67% |
| 2G | Mainly text based services. | 60% |
| Short-range communication | Communication over a few meters or by touch. (Bluetooth, NFC) | 13% |
| WLAN | Mainly indoors or in areas with lot of people. | 1.5% |
| Broad-casting | Sending same information to all cell phones. | 1.1% |
| Satellite | Worldwide access may be needed everywhere including mountains and seas. | 0.47% |
| Combination of two or more access technologies | Non-trivial combination of multiple access technologies. | 0.44% |
| Cell-casting | Sending same information to all cell phones within a base station. | 0.22% |

### 3.2.3 Requirements for Human Technology Interaction (HTI) technologies

Earlier in this paper we discussed the major technical constraints when considering the implementation of mobile service ideas.

Table 5. Human Technology Interaction technologies required in mobile service ideas

| Category | Description | Total |
|---|---|---|
| Image / Video analysis | Extracting information from image or video. | 1.8% |
| Speech synthesis | Ability to produce sound that resembles human speech. | 0.86% |
| Augmented reality | The idea that an observer's experience of an environment can be augmented with computer generated information. | 0.76% |
| Speech recognition | The ability to recognize and carry out voice commands or take dictation. | 0.69% |
| Audio analysis | Extracting information from audio. | 0.49% |
| Gesture recognition | The ability to recognize human gestures, usually hand motion. | 0.22% |
| Gait pattern recognition | The ability to recognize gait. | 0.02% |

HTI-technologies do not seem to be a major obstacle in the track of creating mobile services. However, we know that poor usability and restrictions in HTI-technologies weaken the user experience (Hartmann, Angeli & Sutcliffe 2008).

Developers are investing lots of effort creatingto create solutions thethat make mobile interaction more intuitive. We wanted to look at a number of HTI-technologies in order to see if end users generating service ideas are in need of novel HTI technologies. Table 5 summarizes the analysis.

Based on the analysis we can say that users either can not imagine or do not seem to hunger for services with novel HTI capabilities, such as speech synthesis, augmented reality, speech technologies or gesture recognition.

### 3.2.4 Analysis of Context information required for services ideas users proposed

The use of context information in mobile services has been an active research discipline for already a decade. Researchers and developers have proposed a number of different context-aware mobile services and applications.

Table 6.Context Information required in mobile service ideas

| Category | Description | Total |
|---|---|---|
| Location | Location of the user or some object. | 25% |
| Time | Time that is relevant in a non-trivial way. | 4.7% |
| Activity | What the user is doing. | 1.8% |
| Identities of nearby people | Identities of people that are close-by. | 1.8% |
| Emotion / Mood | How the user or some other party feels. | 1.6% |

Our analysis, summarized in Table 6, confirms the fact that location seems to be the most valuable and versatile context information to be used in mobile services. Even 25% of service ideas require location information. Time (4.7%) is another obvious context information. Identifying the activity of the user, the social context (people nearby) or an emotional situation do not seem to be as important factors for the proposed mobile service ideas.

### 3.2.5 Analysis of needs for integrated or interconnected devices

We also wanted to look at what kind of integrated or interconnected devices people wanted to have in their mobile terminals. The results are summarized in Table 7.

Table 7. Needs for integrated or interconnected devices.

| Category | Description | Total |
|---|---|---|
| Meters/Sensors | Measuring a property. | 6.8% |
| Controllers | Controlling a device. | 3.3% |
| Others | Mobile phone being or using other devices. | 9.9% |

In the analysis, we identified two clear categories, namely Meters/Sensors and Controllers. The first category stands for ideas that have some kind of sensing or measuring functionality. Ideas in this category include, for example, a mobile phone with a step counter or an alcometer. The second category includes ideas related to controlling other devices using a mobile phone, for example, the possibility to open a home door for someone from distance using a mobile phone. In addition to these categories, we found a versatile set of ideas proposing that a mobile phone includes or uses another device, such as a laser pointer or a projector.

## 4 DISCUSSION

Literature claims that people may have unique, interesting and potentially commercially attractive ideas about technologies (Von Hippel 2005). Literature also holds that bright ideas may emerge from inside an organization as well as from the outside (Chesbrough 2003). We challenged these arguments and collected a vast amount of mobile service ideas from students, working-age people, school-children and elderly people. We argue that an in-depth analysis of this material will provide us valuable insight about users' everyday needs and wishes concerning mobile and ubiquitous technologies and services.

In this paper we were particularly interested if user-generated mobile service ideas could be used as a tool in foreseeing the need for technological development. The reasoning behind this idea is as follows: if users would propose needs or ideas that cannot be realised using existing technologies, it would create a potentially attractive target for technology developers.

The discussion is organized around three practical implications of the study. First, we claim that it is relatively simple to collect a large amount of data about people's ideas or needs. The real challenge is to analyze these ideas cost effectively. Second, we claim that the adoption of a vast variety of mobile services is not primarily limited by missing technologies. In fact, the majority of all ideas can be realised with existing technologies. And third, we claim that people are not likely to propose very futuristic service ideas. On the contrary, they are tightly limited to existing understanding of the capabilities of mobile systems. We could say that people are hindered by the technological frame of today.

## 4.1 Consumers freely share their ideas

During the course of the Idea Movement project we have seen that collecting a large amount of idea material is not an overwhelming task. People were willing to participate and reveal their ideas in the brainstorming sessions that we organized in various schools, universities and companies. In these events we offered participants refreshments, such as coffee and snacks, but the participants were not offered any financial or material rewards for participation. This seems to confirm the claims that economic factors are not the primary source of motivation for people who share their ideas (Lüthje 2004). In the beginning of the process of collecting these ideas we thought that IPR and ownership of the ideas would become an important topic of discussion. Surprisingly, only a few people raised the issue. Perhaps this was due to the policy to publish all the ideas, and the fact that these ideas are mainly very short descriptions of a need or service – not detailed business ideas.

Collecting a large amount of ideas is not as difficult as it may sound. Instead (cost) effective processing of these ideas is much greater challenge. We have tried various automated systems for analyzing the data, but they do not seem to work very well. This is mainly due to the free format of ideas. The same idea can be expressed in various ways with different kinds of language, i.e., using standard language, dialect, spoken language or even slang.

An interesting avenue to pursue for processing these ideas is to distribute the workload to large groups of people using the Internet. In this approach persuading and motivating people to work on the ideas requires further consideration. We have done some experiments on carrying out Internet based analysis, but the results fall out of the scope of this paper and thus will be published in the future.

## 4.2 95% of the ideas can be implemented

In our study it became evident that the vast majority of mobile service ideas can be implemented using existing technologies. Technical components of mobile services are typically a mobile terminal (mobile phone) with a browser or a dedicated application, wireless network access and a server with service specific functionality.

According to our analysis only 5% of proposed mobile service ideas had major technical obstacles. One such technical obstacle is, for example, small screen size, which is particularly problematic in navigation and multimedia services. Another technical challenge is limited bandwidth of mobile access. Bandwidth limitation causes troubles in multimedia intensive services such as video-conferencing and multimedia streaming. Power consumption and battery technology may also limit possibilities in few service ideas.

The development of technologies takes some of the current obstacles into history. Some of the new mobile terminals from various vendors have much larger displays than a couple of years ago. Also, new network technologies are provided with improved bandwidth and the computing power will increase, as well as the size of memory.

In fact, we could argue that all proposed mobile service ideas could be realised using existing technology, but the user experience and cost structure might not be quite satisfactory for commercial deployment.

## 4.3 Consumers' technology foresight is limited to existing technologies and paradigms

While we suggested that approximately 95% of all ideas can be realised with existing technologies, it can be said that this method and project produced a large amount of valuable data for companies in quest of new services right now. Even though researchers and developers may be interested in technology forecasts, this material may not readily reveal the future developments of mobile technologies.. Instead, we suggest that extensive pre-processing may enable the discovery of novel patterns not yet identified. We should try to identify some "weak signal" phenomena from the set of ideas instead of stronger trends. These stronger trends tend to be the conventional ideas that are familiar to everyone.

Conservative ideas may be due to the lack of consumers' understanding on the new possibilities of mobile technology or that the majority of users are so tied with their current context that they cannot imagine new paradigms and revolutionary ideas.We are far from being disappointed with the evidently short technology horizon of people. When we consider these ideas from a business point of view, we see that the ideas deemed *Conventional* represent needs of a large group of potential customers that have not been met by the service providers. We see potential breakthrough opportunities by new offerings formed from thousands of raw ideas.

We were also expecting to see more ideas related to new developments in the Internet such as social networking and the creation of mobile user-generated content. Another topic that we expected to see more was commerce, payment and identification of users using mobile technology. Both of these developments are active in the Internet, but participants of these events did not see the need for carrying out these tasks with their mobile systems.

Many of these services, however, require that new revenue-models should be defined, which would enable the penetration of mobile services based on the content and interest rather than merely on data traffic costs.

## 5 CONCLUSIONS AND FUTURE WORK

The 40,000 ideas for new mobile services expressed by the participants are far from well-defined business concepts ripe for commercialization. The ideas still require further development and professional elaboration before they can be introduced to the market as products or services.

In this study we presented the findings based on a technical analysis of 4,000 ideas. The findings were grouped into three principal practical implications: 1) Consumers share their ideas freely, 2) vast majority of the ideas can be implemented by means of existing technologies, and 3) consumers' technology foresight appears to be restricted by their experience with current technology and paradigms.

In order to advance this development we plan to carry out a set of analyses for the idea database. One particular topic that we are interested in is comparing the differences in ideas of people of different age. We want to explore this further in future studies. We will continue to expand the Idea Movement to other countries, cultures and nationalities in the near future. The Idea Movement's (www.idealiike.fi) idea bank is now open in Finnish for all commercial, research and educational purposes. The ideas will be opened up in English as well. This enables the development of these ideas in global context.

We predict that the actual value of these ideas is generated in the idea refinement phase where hundreds of ideas are combined and enriched through the conceptualization process. This is a task reserved for companies and organizations with the ability to efficiently commercialize the services. Idea Movement, therefore, makes a link between the intellectual capital of ordinary citizens and the technology business know-how of companies which results in a potential win-win situation.

## 6 ACKNOWLEDGEMENTS

We acknowledge Idea Movement Partners, Maaretta Törrö and Anssi Öörni.